\documentclass[%
 aps,
 amsmath,amssymb,
 reprint,%
floatfix,
]{revtex4-2}

\usepackage{graphicx}
\usepackage{dcolumn}
\usepackage{bm}
\usepackage{etoolbox}
\usepackage[inline]{enumitem}
\setcitestyle{numbers}

\makeatletter
\def\@email#1#2{%
 \endgroup
 \patchcmd{\titleblock@produce}
  {\frontmatter@RRAPformat}
  {\frontmatter@RRAPformat{\produce@RRAP{*#1\href{mailto:#2}{#2}}}\frontmatter@RRAPformat}
  {}{}
}%
\makeatother
\begin{document}
\title{Striking Similarities in Dynamics and Vibrations of  2D Quasicrystals and Supercooled Liquids}
\author{Edwin A. Bedolla-Montiel}
\affiliation{Soft Condensed Matter \& Biophysics, Debye Institute for Nanomaterials Science, Utrecht University, Princetonplein 1, 3584 CC Utrecht, Netherlands.}

\author{Marjolein Dijkstra}
\affiliation{Soft Condensed Matter \& Biophysics, Debye Institute for Nanomaterials Science, Utrecht University, Princetonplein 1, 3584 CC Utrecht, Netherlands.}
\email{m.dijkstra@uu.nl, e.a.bedollamontiel@uu.nl}

\date{\today}

\begin{abstract}
	We investigate the interplay between structure and dynamics in two structurally distinct two-dimensional systems: a dodecagonal quasicrystal (DDQC) and a supercooled binary liquid. Using molecular dynamics simulations, we uncover striking dynamical similarities despite their fundamentally different structural organizations. 
    Both systems exhibit pronounced dynamic heterogeneities, as evidenced by the cage-trapping plateaus in the mean-squared displacement and the pronounced peaks in the non-Gaussian parameter. In both cases, we observe a strong  correlation between local structural order and dynamic propensity, indicating similar structure-dynamics relationships, albeit driven by distinct microscopic mechanisms.
	Despite these parallels, their vibrational properties diverge: the DDQC exhibits multiple peaks linked to phason dynamics, while the supercooled liquid displays a characteristic boson peak. Analysis of vibrational eigenmodes shows that both systems exhibit extended modes at low frequencies. At high frequencies, however, the DDQC maintains a higher density of topological defects, reflecting its quasi-long-range order. Finally, we contextualize these findings by comparing both systems to a square crystal. While the dynamics appears similar across all three systems, the vibrational and topological features clearly distinguish the DDQC and glass from the crystalline state. These results underscore a surprising universality in dynamical behavior across structurally diverse systems and provide new insights into how structural organization shapes motion in soft-matter systems.
\end{abstract}

\maketitle

\section{\label{sec:intro}Introduction}
At the intersection of ordered and disordered matter lie two captivating states of matter: quasicrystals (QCs), with their nonperiodic but long-range order, and supercooled liquids approaching the glass transition.
Despite their fundamentally different structural organizations, these systems exhibit striking dynamical similarities that challenge our traditional understanding on the relationship between structure and dynamics.
This resonates with the work of Angel  ~\cite{angellAmorphousStateEquivalent2000}, who demonstrated that amorphous materials can undergo transitions analogous to (first-order) crystallization, resulting in  distinct glassy states with unique thermodynamic signatures.
He related this behavior to QCs, whose nonperiodic, unconventional  ordering  blurs the boundary between disordered and crystalline phases. 

Quasicrystals exhibit long-range orientational order without translational periodicity. First  discovered in metallic alloys by Shechtman {\em et al.}~\cite{shechtmanMetallicPhaseLongRange1984},  QCs have since attracted growing interest in soft matter, particularly following their experimental and computational realization in colloidal, polymeric, and molecular systems~\cite{zengSupramolecularDendriticLiquid2004,hayashidaPolymericQuasicrystalMesoscopic2007,forsterQuasicrystallineStructureFormation2013,urgelQuasicrystallinityExpressedTwodimensional2016,liuRationalDesignSelfAssembly2019,forsterQuasicrystalsTheirApproximants2020,noya2021design,schenk2DHoneycombTransformation2022,liGrowthQuasicrystalrelatedStructure2023,zengColumnarLiquidQuasicrystal2023,fayenSelfassemblyDodecagonalOctagonal2023a,platiQuasicrystallineOrderVibrating2024,zhou2024colloidal}. The principal interest in these structures arises from their unique physical properties and potential applications, particularly in photonics, where their aperiodic order enables unique transmission characteristics, band gaps, and disorder-enhanced transport phenomena~\cite{matsuiTransmissionResonancesAperiodic2007,leviDisorderEnhancedTransportPhotonic2011,vardenyOpticsPhotonicQuasicrystals2013,yuEngineeredDisorderPhotonics2021}.

Supercooled liquids and glasses have long been central to the study of complex condensed matter dynamics. These systems exhibit hallmark phenomena such as dynamical heterogeneity~\cite{kegelDirectObservationDynamical2000,berthierDynamicHeterogeneityAmorphous2011}, transient caging~\cite{doliwaCageEffectLocal1998,weeksPropertiesCageRearrangements2002,hunterPhysicsColloidalGlass2012}, and cooperative particle  rearrangements~\cite{zhangStringlikeCooperativeMotion2013,zhangRoleStringlikeCollective2015,ishinoMicroscopicStructuralOrigin2025}. 
As temperature decreases, supercooled liquids exhibit a pronounced dynamical  slowdown that eventually leads to the glass transition, where molecular motion becomes severely constrained in the absence of crystallization~\cite{janssenModeCouplingTheoryGlass2018}. Unlike crystals or quasicrystals, their structural complexity emerges not from long-range order but from disordered local rearrangements and medium-range correlations that strongly influence the dynamics.

Despite extensive research on QCs and supercooled liquids as separate systems, comparative studies that explore their structural and dynamical similarities remain limited. Recent work by Cao {\em et al.}~\cite{caoPhononDynamics3D2025}  demonstrated that three-dimensional icosahedral QCs and supercooled liquids exhibit comparable phonon dynamics. However, a systematic comparison of structure-dynamics relationships, vibrational properties, and topological features in two-dimensional (2D) systems has yet to be undertaken. The apparent paradox that structurally distinct systems can display remarkably similar  dynamical behavior motivates the present study. We focus on 2D systems, which offer distinct advantages: enhanced accessibility for visualizing  structural motifs and collective motions, and direct relevance to  experimental colloidal systems where such behaviors can be observed in real time and real space~\cite{vaibhavExperimentalIdentificationTopological2025}.

Previous studies have extensively examined structure-dynamics relationships in both glasses and quasicrystals, though largely in isolation. In glasses, dynamical heterogeneity has been closely linked to  local structural ordering, with particular emphasis on identifying structural motifs that correlate with particle  mobility~\cite{tongRevealingHiddenStructural2018,paretAssessingStructuralHeterogeneity2020a}. Medium-range structural features have also been associated with  cooperative rearrangements and string-like motion~\cite{zhangStringlikeCooperativeMotion2013,zhangRoleStringlikeCollective2015}. In QCs, research has primarily focused on phason dynamics~\cite{socolar1986phonons,a2013phason}, particle flips~\cite{engelDynamicsParticleFlips2010}, and self-diffusion mechanisms~\cite{kaluginMechanismSelfDiffusionQuasiCrystals1993,zhaoAtomisticMechanismsDynamics2025}. Notably, Zhao {\em et al.}~\cite{zhaoAtomisticMechanismsDynamics2025} proposed that QCs may represent an intermediate state between crystals and glasses in terms of their dynamic properties, although their focus was  on diffusion rather than vibrational or topological characteristics. Beyond a recent comparative study in  three-dimensional systems~\cite{caoPhononDynamics3D2025},  systematic investigations of  dynamical heterogeneity, vibrational properties, and associated  topological features in two-dimensional QCs and glasses have received limited attention.

In this work, we employ molecular dynamics simulations to systematically compare the structural and dynamical properties of a two-dimensional dodecagonal quasicrystal (DDQC) and a binary supercooled liquid. The DDQC is stabilized using a continuous square-shoulder-like potential, specifically tuned to stabilize dodecagonal symmetry, while the supercooled liquid is modeled via a binary mixture interacting through  a soft repulsive pair potential.  To mitigate  the challenges posed by Mermin--Wagner fluctuations inherent to two-dimensional systems, we implement cage-relative analysis techniques for all dynamical observables~\cite{illingMerminWagnerFluctuations2017}. We quantify dynamical heterogeneity using the non-Gaussian parameter, and establish structure-dynamics correlations via dynamic propensity maps, which we compare against local structural order parameters:  orientational order in the DDQC and excess entropy in the supercooled liquid. We further investigate  vibrational properties by computing the dynamical matrix of energy-minimized configurations, extracting the vibrational density of states, participation ratios, and topological defect distributions  of the vibrational eigenmodes. Finally, we compare the quasicrystal and  supercooled liquid to a square crystal. While the dynamics of all three systems appear similar, their vibrational properties provide a distinct fingerprint that differentiates them.

Our analysis uncovers both striking similarities and key differences  between QCs and supercooled liquids,  reflecting the interplay between  their distinct structural organizations and dynamic behavior. Both systems exhibit cage-trapping plateaus in their mean-squared displacement and pronounced peaks in their non-Gaussian parameters, indicative of comparable cage-escape mechanisms despite their differing structural arrangements. In both cases, we observe a correlation between local structural order and dynamic propensity---a relationship well-established in glass-forming liquids but less explored in QCs.
In contrast, their vibrational properties reveal substantial differences: the DDQC exhibits multiple peaks associated with phason modes, consistent with its quasiperiodic order, while the supercooled liquid features a characteristic boson peak. Analysis of topological defects in the vibrational eigenmodes demonstrates different high-frequency behavior: the QCs retains a higher density of defects, underscoring the persistence of  quasi-long-range structural correlations. Finally, we contextualize our findings by comparing the results obtained to those of a square crystal, revealing that while both the DDQC and supercooled liquid exhibit qualitatively similar dynamical behavior, they diverge significantly from the crystal in terms of vibrational properties and topological features.

The paper is organized as follows. In Section~\ref{sec:methods}, we describe the two model systems  and  the simulation details. We analyze the dynamical heterogeneity and structure-dynamics correlations in Section~\ref{sec:dynamics}. 
Section~\ref{sec:vibrational} examines the vibrational properties through the density of states, participation ratios, and topological defect distributions. In   Section~\ref{sec:crystal-similarities},  the dynamical and vibrational properties of the dodecagonal quasicrystal and the supercooled liquid are compared to a square crystal.
Finally, Section~\ref{sec:conclusions} summarizes  our findings and offers an outlook.

\begin{figure}
	\includegraphics[scale=0.95]{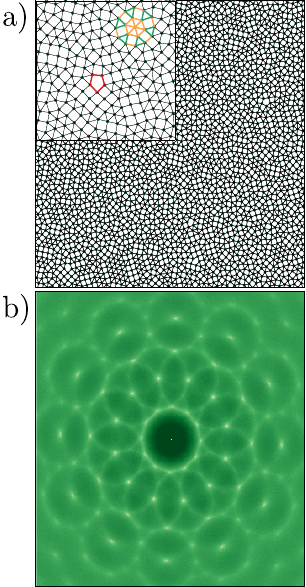}
	\caption[diff]{
		\begin{enumerate*}[label=\textbf{\alph*})]
            \item Typical configuration of a two-dimensional dodecagonal quasicrystal  of  \(N=4096\) particles. Interparticle bonds (black lines) are drawn for particle distances  \(r < 1.4 \sigma \). The inset highlights a characteristic dodecagonal "wheel" motif in orange and green---an essential motif  of a true quasicrystal formed with a random square-triangle tiling~\cite{oxborrowRandomSquaretriangleTilings1993,imperor-clercSquaretriangleTilingsInfinite2021}. The inset also shows in red a defect structure, frequently observed in soft-matter quasicrystals, and recently shown to enhance thermodynamic stability~\cite{ulugolDefectsEnhanceStability2024}.\label{fig:diff-qc-snapshot}
            \item Intensity map of the static structure factor \(S({\bm{k}})\)  corresponding to the dodecagonal quasicrystal configuration, revealing clear 12-fold rotational symmetry characteristic of dodecagonal order.\label{fig:diff-qc}
		\end{enumerate*}}
	\label{fig:diffraction}
\end{figure}

\section{\label{sec:methods} Model Systems and Simulation Details}

We performed molecular dynamics (MD) simulations on a two-dimensional dodecagonal quasicrystal (DDQC) and a supercooled  binary liquid mixture   using the LAMMPS software package~\cite{THOMPSON2022108171,plimpton_2024_10806852}.
Simulations were carried out in the canonical (\(NVT\)) ensemble  for systems containing \(N=4096\), \(N=10976\) or \(N=16384\) particles.
Smaller system sizes were employed primarily for computing dynamical observables such as the mean-squared displacement, while larger systems were used for structural and vibrational analyses.
Periodic boundary conditions were applied in all spatial directions. 

All dynamical quantities were measured after equilibrating the system in the \(NVT\) ensemble until both  temperature and total energy were constant. Following equilibration, all thermostats were removed and the  simulations were continued in the \(NVE\) ensemble for data collection.

\subsection{2D Dodecagonal Quasicrystal}
To model a two-dimensional dodecagonal quasicrystal (DDQC), we employ  a continuous interaction potential designed to mimic the behavior of a square-shoulder potential. The potential is given by
\begin{equation}
	u_{\mathrm{QC}}(r) / \epsilon = {\left( \frac{\sigma}{r} \right)}^{14} + \frac{1 - \tanh{\left[ k(r - \delta)\right]}}{2} \, ,
	\label{eq:qc12}
\end{equation}
where \(\epsilon\) sets the energy scale, \(\sigma\) represents the typical core diameter, \(k \sigma=10\) controls the steepness of the repulsive shoulder, and \(\delta=1.35 \sigma\) determines the interaction  range. The parameters \(k\) and \(\delta\) are carefully tuned to stabilize a DDQC phase~\cite{padillaPhaseBehaviorTwodimensional2020,coliInverseDesignSoft2021}; varying these values can yield QCs with different orientational symmetries.  All simulations were performed at a fixed number density of \(\rho \sigma^{2} = 0.935\), selected based on previously reported phase diagrams~\cite{padillaPhaseBehaviorTwodimensional2020}, and the control parameter is the dimensionless temperature \(k_{B} T / \epsilon\), where \(T\) is the absolute temperature and \(k_{B}\) denotes Boltzmann's constant. The stability region for  the DDQC is very narrow, a feature also demonstrated for  a binary DDQC mixture~\cite{fayenQuasicrystalBinaryHard2024}. We performed MD simulations on the DDQC using an integration time step of \(\Delta t = 0.001 \tau_{\mathrm{QC}}\), where  \(\tau_{\mathrm{QC}}=\sigma \sqrt{m/\epsilon}\) denotes the MD time unit with $m$ the particle mass.
Temperature control was implemented using  the Bussi-Donadio-Parrinello thermostat~\cite{bussiCanonicalSamplingVelocity2007} with a damping parameter of \(0.1 \tau_{\mathrm{QC}}\).

The DDQC was generated by following a careful cooling protocol. Initially, the system was equilibrated at a high temperature of \(k_{B} T / \epsilon = 1\) for at least \(10^6\) MD time steps. Subsequently, the temperature was reduced using a linear cooling rate over \(10^7\) MD time steps to reach a target temperature range \(k_B T/\epsilon \in [0.15, 0.18]\), within which the DDQC is thermodynamically stable. Once the target temperature was reached, the DDQC was further equilibrated for at least \(10^8\) MD time steps to ensure the formation and stability of the quasicrystalline structure.

An exemplary configuration of the DDQC is shown in Fig.~\ref{fig:diffraction}\ref{fig:diff-qc-snapshot}, where the structure exhibits the characteristic random square-triangle  tiling that constitutes the typical motifs of the quasicrystalline structure. The inset  highlights  the  dodecagonal wheel, a hallmark structural unit of square-triangle quasicrystals, commonly observed in all true QCs formed from the square-triangle random tiling~\cite{oxborrowRandomSquaretriangleTilings1993,imperor-clercSquaretriangleTilingsInfinite2021}. This feature distinguishes the DDQC  from structurally related but distinct phases, such as the \(\Sigma\) phase~\cite{imperor-clercSquaretriangleTilingsInfinite2021}. Figure~\ref{fig:diffraction}\ref{fig:diff-qc} shows the corresponding intensity map of the static structure factor, revealing the expected twelve-fold rotational symmetry. The peaks are not perfectly aligned due to residual phason strain and the presence of defects; such imperfections can vary across simulations and may be reduced by energy-minimization procedures.

\begin{figure}
	\includegraphics[scale=0.95]{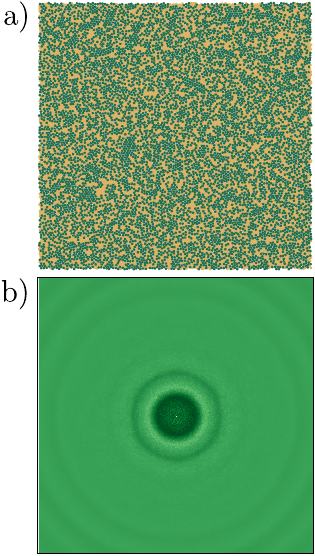}
	\caption[diff-glass]{
		\begin{enumerate*}[label=\textbf{\alph*})]
			\item Typical configuration of a two-dimensional supercooled binary liquid mixture of \(N=4096\) particles. The two species are distinguished by color: green denotes the larger particles, and yellow the small ones.
            \item Intensity map of the static structure factor \(S({\bm{k}})\)  for the supercooled liquid, showing diffuse concentric rings indicative of the absence of long-range translational order.
		\end{enumerate*}}
	\label{fig:diffraction-glass}
\end{figure}

\subsection{2D Supercooled Binary Liquid Mixture}
For the supercooled liquid, we employ an equimolar binary mixture of particles with diameters \(\sigma_{\mathrm{S}}\) and \(\sigma_{\mathrm{L}}\), interacting via a Weeks-Chandler-Andersen  potential. This model is widely used due to its strong frustration against crystallization~\cite{hamanakaTransitionsCrystalGlass2006a,tanakaCriticallikeBehaviourGlassforming2010,wuTopologyVibrationalModes2023}. The  continuous nature of this potential makes it particularly well suited for analyzing vibrational properties, in contrast to discontinuous potentials such as the hard-sphere potential, which require specialized  techniques for such analyses~\cite{arceriVibrationalPropertiesHard2020}.
The interaction potential is defined as
\begin{equation}
	u_{\mathrm{LS}}(r) = 4 \epsilon \left[ {\left(\frac{\sigma_{\mathrm{LS}}}{r}\right)}^{12} - {\left(\frac{\sigma_{\mathrm{LS}}}{r}\right)}^{6} \right] + \epsilon  \, ,
	\label{eq:lennard-jones}
\end{equation}
where \(\sigma_{\mathrm{LS}} = (\sigma_{\mathrm{L}}+\sigma_{\mathrm{S}})/2\) represents the cross-interaction term, making the mixture additive. The constant \(\epsilon\) ensures that the interaction potential  smoothly vanishes at the cutoff radius,  \(u_{\mathrm{LS}}(r_{\mathrm{cut}}) = 0\), with \(r_{\mathrm{cut}}=2^{1/6}\sigma_{\mathrm{LS}}\), making the potential purely repulsive and continuous.
The size ratio is set to \(\sigma_{\mathrm{L}} / \sigma_{\mathrm{S}} = \sqrt{2}\) to introduce size disparity and suppress crystallization~\cite{hamanakaTransitionsCrystalGlass2006a,kawasakiStructuralSignatureSlow2011}.
The particle masses follow  \(m_{\mathrm{L}} / m_{\mathrm{S}}={\left(\sigma_{\mathrm{L}} / \sigma_{\mathrm{S}}\right)}^{2}\).
Similar to the DDQC system, the control parameter  is the reduced temperature \(k_{B} T / \epsilon\), and all simulations are performed at a fixed number density of \(\rho \sigma^{2} = 0.75\).

We performed MD simulations on the supercooled binary liquid mixture using an integration time step of \(\Delta t = 0.001 \tau_{\mathrm{G}}\), where  \(\tau_{\mathrm{G}}=\sigma_{\mathrm{L}} \sqrt{m_{\mathrm{L}} / \epsilon}\) represents the MD  time unit for the glass. We employ the Bussi-Donadio-Parrinello thermostat to keep the temperature fixed~\cite{bussiCanonicalSamplingVelocity2007} using a damping parameter of \(0.1 \, \tau_{\mathrm{G}}\).

The supercooled binary liquid mixture was generated using an  annealing protocol similar to that of the DDQC. The system was first equilibrated at a high temperature of \(k_{B} T / \epsilon = 5\) for \(10^6\) MD time steps,  then linearly cooled over \(10^7\) time steps to a target temperature \(k_BT/\epsilon \in [0.1, 1]\).  At the target temperature, the system was equilibrated for at least \(10^8\) MD time steps to ensure thermal and structural stability.

Figure~\ref{fig:diffraction-glass} shows a typical configuration of the supercooled liquid, along with the corresponding intensity map of the static structure factor. Unlike the sharp Bragg peaks seen for the DDQC, the structure factor here displays diffuse concentric rings,  consistent with the absence  of long-range structural order characteristic of amorphous systems.

\section{\label{sec:dynamics} Dynamical Heterogeneity and Structure–Dynamics Correlation}

\subsection{\label{sec:msd} Mean-squared displacement}
The mean-squared displacement (MSD) is a fundamental measure for characterizing particle dynamics in condensed matter systems. However, in two-dimensional systems, direct measurements of dynamical quantities are  complicated by long-wavelength thermal fluctuations in particle positions, which  arise as a consequence of the Mermin-Wagner theorem~\cite{merminAbsenceFerromagnetismAntiferromagnetism1966}. These fluctuations can obscure true signatures of dynamic arrest or caging. To mitigate this effect, we adopt the cage-relative (CR) MSD approach, as   proposed in Ref.~\cite{illingMerminWagnerFluctuations2017}, which filters our collective motions by redefining displacements relative to the local neighborhood. The CR-MSD is computed as
\begin{widetext}
	\begin{equation}
        {\langle r^{2}(t;t_{0}) \rangle}_{\mathrm{CR}} =
		\frac{1}{N} \left\langle \sum_{i=1}^{N}\left[
		(\bm{r}_{i}(t + t_{0}) - \bm{r}_{i}(t_{0})) -
        \frac{1}{N_{i}}\sum_{j=1}^{N_{i}}(\bm{r}_{j}(t+t_{0}) - \bm{r}_{j}(t_{0}))
		\right]^2\right\rangle \, .
		\label{eq:cr-msd}
	\end{equation}
\end{widetext}
In Eq.~\eqref{eq:cr-msd}, the second term  subtracts the displacement of the center of mass of the local cage surrounding each particle \(i\), thereby isolating the motion of the particle relative to its neighbors. The cage is defined by the \(N_{i}\) nearest neighbors of particle \(i\),  identified using a fixed cutoff radius \(r_{c}\). We use  \(r_{c}=1.4 \, \sigma\) for the DDQC and \(r_{c}=1.2 \, \sigma_{L}\) for the supercooled liquid.

\begin{figure}
	\includegraphics[width=0.85\columnwidth]{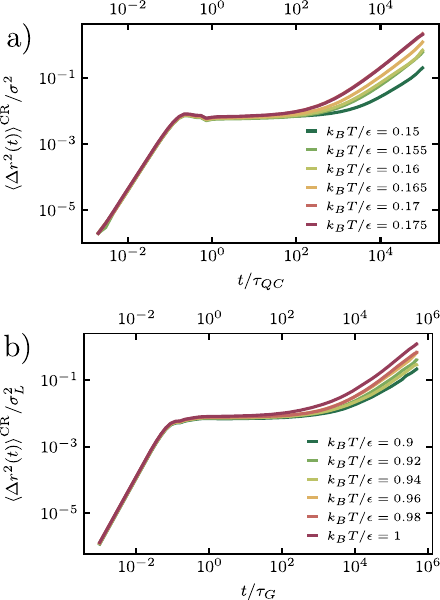}
	\caption[msd]{Cage-relative mean-squared displacement (MSD) of 
		\begin{enumerate*}[label=\textbf{\alph*})]
			\item the dodecagonal quasicrystal, and \label{fig:msd-qc}
			\item the binary supercooled liquid, as a function of time for varying temperatures. \label{fig:msd-glass}
		\end{enumerate*}
        The emergence of a cage-trapping plateau in both systems indicates transient localization of particles within cages formed by their neighbors, before transitioning to diffusive motion at longer times.}
	\label{fig:msd}
\end{figure}

We computed the cage-relative MSD for both the DDQC and the supercooled liquid at various temperatures, as presented in Fig.~\ref{fig:msd}.
Following an initial ballistic regime, both systems exhibit a pronounced cage-trapping plateau in the MSD, indicative of particle localization within transient cages.
The notable similarity observed between the two systems highlights analogous intermediate-time dynamics, wherein particles remain temporarily trapped in their local environments before escaping their cages and transitioning into diffusive motion.
Similar plateaus have also been observed in crystal and hexatic phases~\cite{meerDynamicalHeterogeneitiesDefects2015a}, where collective rearrangements involving loops of particles and the diffusion of vacancy-interstitial pairs manifest explicitly as dynamical heterogeneity.
We explore this aspect further in Section~\ref{sec:crystal-similarities}.

The MSD plateau persists for comparable durations in both systems, corresponding to a regime where particle mobility is significantly restricted. 
In supercooled liquids, this plateau reflects the formation of transient cages that hinder local particle dynamics, a phenomenon  well-established in experiments and simulations~\cite{angellTenQuestionsGlassformers2000,tongRevealingHiddenStructural2018}.
In contrast, the trapping mechanism in the DDQC is somewhat distinct: particle rearrangements require overcoming local energy barriers associated with the quasicrystalline potential energy landscape.
This mechanism, initially predicted theoretically~\cite{kaluginMechanismSelfDiffusionQuasiCrystals1993}, has since been corroborated  by simulation studies~\cite{engelDynamicsParticleFlips2010}.
At sufficiently long timescales, once particles escape their  transient cages, both systems exhibit diffusive behavior, provided thermal equilibrium is maintained.
As expected, the long-time self-diffusion coefficient increases with temperature, as seen in the increased slope in the long-time regime of the MSD. 
However, the microscopic diffusion mechanisms differ: in the DDQC, diffusion is facilitated by collective particle rearrangements associated with local tile rearrangements, enabling particles to travel large distances.
In contrast, diffusion in the  supercooled liquid  emerges from collective motions, such as string-like particle rearrangements~\cite{zhangStringlikeCooperativeMotion2013,zhangRoleStringlikeCollective2015}, allowing particles to diffuse extensively through cooperative dynamics.

\subsection{\label{sec:alfa2} Non-Gaussian parameter}
While the MSD analysis reveals plateau regimes indicative of transient particle caging,  it does not fully characterize the heterogeneous nature of particle dynamics. In systems with dynamical heterogeneity, particle displacements, \(\Delta r(t) = \sqrt{({\bf r}(t+t_0)-{\bf r}(t_0))^2}\), deviate significantly from the Gaussian distribution expected for normal diffusive motion~\cite{berthierDynamicHeterogeneityAmorphous2011}. To quantify these deviations, we employ the non-Gaussian parameter, \(\alpha_{2}(t)\), which measures the extent to which the distribution of particle displacements differs from Gaussian behavior. For our two-dimensional systems, we evaluate \(\alpha_{2}(t)\) using the CR coordinate system introduced in Section \ref{sec:msd}
\begin{equation}
	{\langle \alpha_{2}(t) \rangle}_{\mathrm{CR}} =
	\frac{{\langle \Delta r^{4}(t) \rangle}_{\mathrm{CR}}}{2\,{\langle \Delta r^{2}(t) \rangle}_{\mathrm{CR}}^{2}} - 1 \, ,
	\label{eq:cr-alfa2}
\end{equation}
where \(\langle \cdots \rangle \) denotes an ensemble average over all particles and initial times $t_0$. In the case of purely Gaussian dynamics, \(\alpha_{2}(t) = 0\), while positive values indicate broader tails in the displacement distribution, a signature of dynamical heterogeneity.

\begin{figure}
	\includegraphics[width=0.85\columnwidth]{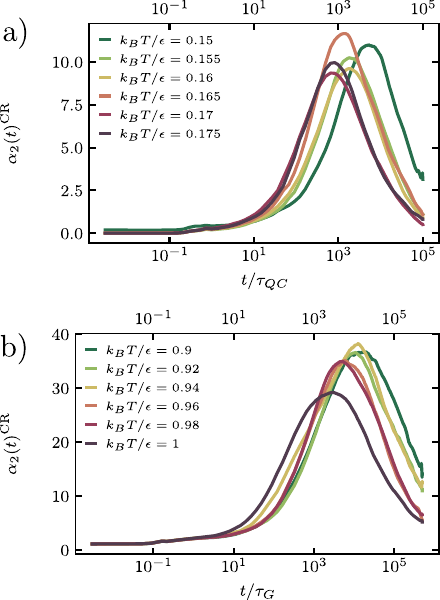}
	\caption[a2]{Cage-relative non-Gaussian parameter \(\alpha_{2}(t)\) of
    \begin{enumerate*}[label=\textbf{\alph*})]
        \item the dodecagonal quasicrystal, and \label{fig:alfa2-qc}
        \item the supercooled binary liquid mixture, as a function of time for varying temperatures. \label{fig:alfa2-glass}
    \end{enumerate*}\ignorespaces
    Peaks in \(\alpha_{2}(t)\) highlight the emergence of dynamical heterogeneity, with characteristic time scales \(t^{*}\) associated with cage-breaking events.}
	\label{fig:alfa2}
\end{figure}

Figure~\ref{fig:alfa2} shows the non-Gaussian parameter \(\alpha_{2}(t)\) for  the  DDQC and the supercooled liquid at various temperatures.
Both systems exhibit  pronounced peaks in \(\alpha_{2}(t)\), indicating clear  dynamical heterogeneity.
These peaks occur at characteristic times \(t^{*}\)  corresponding to the cage-breaking regime when some particles escape their local cages while others remain trapped.
The peak height reflects the degree of dynamical heterogeneity, with higher peaks indicating greater variations in particle mobility across the system.

A notable similarity emerges between the two systems: the DDQC exhibits non-Gaussian behavior remarkably similar to that of the supercooled liquid, with comparable peak magnitudes and characteristic time scales despite their fundamentally different structural organization.
However, the temperature dependence of these features differs in a subtle but important way.
In the supercooled liquid (Fig.~\ref{fig:alfa2}\ref{fig:alfa2-glass}), the characteristic time \(t^{*}\) decreases monotonically with increasing temperature, consistent with thermally activated cage-escape processes. In contrast, the DDQC (Fig.~\ref{fig:alfa2}\ref{fig:alfa2-qc}) displays a non-monotonic temperature dependence,  suggesting that particle rearrangements are governed by  competing effects. Notably, at \(k_{B}T / \epsilon=0.155\), the characteristic time \(t^{*}\) is slightly larger than at \(k_{B}T / \epsilon=0.16\).

This non-monotonic behavior differs from previous observations in DDQCs reported in Ref.~\cite{zhaoAtomisticMechanismsDynamics2025}, where the non-Gaussian parameter decreased monotonically with temperature. Our results suggest that the temperature dependence of dynamical heterogeneity in QCs may be more complex than previously thought and could depend sensitively on the specific temperature range examined. The observed non-monotonic behavior may arise from competing mechanisms in QC dynamics: while higher temperatures generally enhance particle mobility, they may also destabilize local structural motifs that govern rearrangement processes, resulting in complex and nontrivial effects on particle motion.
Additionally, it might be that the density of defects that are already in the system when exploring the different temperature range might alter the behavior of the dynamical properties.

\subsection{Correlation Between Dynamic Propensity and Local Order in Dodecagonal Quasicrystals}

These findings suggest that dynamical heterogeneity in the DDQC arises from  structural competition at the local level, likely involving the rearrangement of quasicrystalline motifs. To investigate this further, we analyze the  correlation between dynamic propensity and local structural order. 
The dynamic propensity of a particle quantifies its future mobility as determined by its local environment. It is defined as the absolute  displacement of each particle over a given  time interval, averaged over many  trajectories  initiated from the same  configuration but with randomized velocities. This metric captures the  component of  dynamical heterogeneity  encoded
in the structure.

To this end, we perform \(M=50\) independent simulations starting from the same initial configuration, but with velocities randomly drawn from a Maxwell-Boltzmann distribution.
The dynamic propensity of particle \(i\) is then computed as the average absolute displacement with cage-relative coordinates, defined as
\begin{equation}
        {\langle d_{i}(t) \rangle}_{\mathrm{CR}} = \left\langle
        \left\lvert \bm{r}_{i}(t) - \bm{r}_{i}(0)  -
        \frac{1}{N_{i}}\sum_{j=1}^{N_{i}}  \bm{r}_{j}(t) - \bm{r}_{j}(0) \right\rvert
        \right\rangle \, .
    \label{eq:cr-dynamic-propensity}
\end{equation}
over a timescale  \(t = 10^{6} \tau_{QC}\), and the average \(\left\langle \cdots \right\rangle\)is taken over the \(M\) replicas.

\begin{figure*}
	\includegraphics[width=0.95\textwidth]{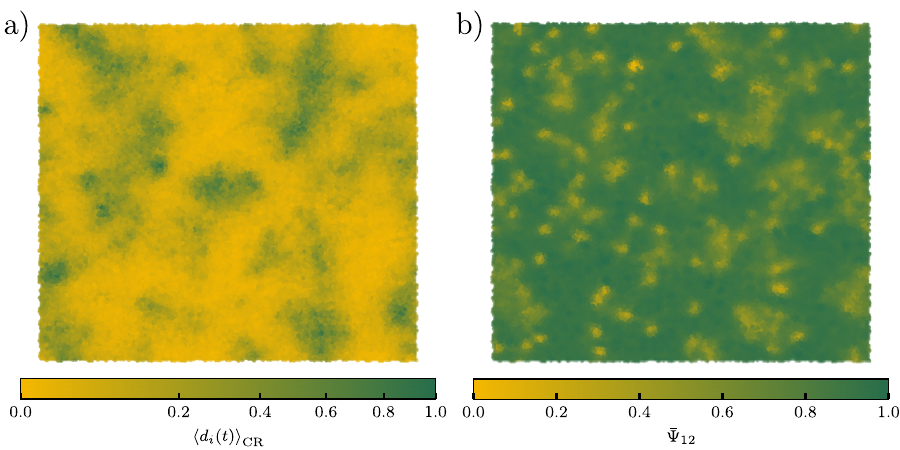}
	\caption[corr-qc]{
		\begin{enumerate*}[label=\textbf{\alph*})]
            \item Dynamic propensity with cage-relative coordinates \(\langle d_{i} (t) \rangle_{\mathrm{CR}}\) \label{fig:qc-propensity}
            and \item local orientational order parameter \(\bar{\Psi}_{12}(i)\) for a dodecagonal quasicrystal with \(N = 16384\) particles at temperature \(k_{B} T / \epsilon = 0.17\) and density \(\rho \sigma^{2} = 0.935\), evaluated over a time interval  \(t = 10^{6} \tau_{QC} .\) We use \(M=50\) replicas of the same initial configuration with velocities randomly sampled  from a Maxwell-Boltzmann distribution, to compute the average  absolute displacement across all replicas. 
		\label{fig:qc-structure}
		\end{enumerate*}
	}
	\label{fig:corr-qc}
\end{figure*}

In Fig.~\ref{fig:corr-qc}\ref{fig:qc-propensity}, we show the dynamic propensity map for the DDQC with \(N=16384\) particles at  temperature $k_BT/\epsilon=0.17$ and density \(\rho\sigma^2=0.935\). 
Pronounced dynamical heterogeneity is clearly visible, with distinct regions of high
and low particle mobility persisting over the analyzed time interval.
This dynamical  heterogeneity aligns well with the peaks observed in the non-Gaussian parameter (Fig.~\ref{fig:alfa2}\ref{fig:alfa2-qc}), confirming that particles in the DDQC experience significantly different local environments that influence their mobility.
To quantify the local structural order in the DDQC, we propose a similar approach as shown in Ref.~\cite{boattini2020autonomously,bedollamontiel2025relationshipstructuredynamicsicosahedral} to use the first and second shell of neighbors to obtain the local orientational order parameter \(\psi_{n}(i)\).
First, we define for any particle \(i\) the complex quantities
\begin{equation}
	\psi_{n} (i) = \frac{1}{\mathcal{N}_{b}(i)} \sum_{j \in \mathcal{N}_{b}(i)} e^{\imath \, n\, \theta_{ij}} \, ,
\end{equation}
with \(n \in \mathbb{Z}\) selected to probe an \(n\)-fold angular symmetry.
We also have that \(\imath = \sqrt{-1}\), \(\mathcal{N}_{b}(i)\) is the number of nearest neighbors of particle \(i\), and \(\theta_{ij}\) denotes the angle between the vector \(\bm{r}_{ij} = \bm{r}_{j} - \bm{r}_{i}\) and the \(x\)-axis.
Neighbors were identified using a cutoff radius of \(r_{c} = 1.4 \sigma\).
In two dimensions, we can define a set of rotationally invariant quantities using the modulo,
\begin{equation}
	\left\lvert \psi_{n} (i) \right\rvert = \left[ \Re{\psi_{n} (i)}^{2} + \Im{\psi_{n} (i)}^{2} \right]^{1/2} \, .
\end{equation}
Finally, we average over the second-neighbor shell as follows,
\begin{equation}
	\bar{\psi}_{n} (i) = \frac{1}{\mathcal{N}_{b}(i) + 1} \left[ \left\lvert \psi_{n} (i) \right\rvert + \sum_{k \in \mathcal{N}_{b}(i)} \left\lvert \psi_{n} (k) \right\rvert \right] \, .
	\label{eq:psi}
\end{equation}
Following Ref.~\cite{tanakaCriticallikeBehaviourGlassforming2010}, we computed the time-averaged orientational order parameter for each particle \(i\) as
\begin{equation}
	\bar{\Psi}_{n}(i) = \frac{1}{\tau_{d}} \int_{t'}^{t'+\tau_{d}} \bar{\psi}_{n} (i) \, dt,
	\label{eq:psi-time}
\end{equation}
We evaluated \(\bar{\Psi}_{n}(i)\) over a total duration of \(\tau_{d} = 10^{6} \tau_{QC}\), for symmetry values \(n = 12\), corresponding to the primary twelve-fold symmetry of the DDQC. 

In Fig.~\ref{fig:corr-qc}\ref{fig:qc-structure}, we display the combined structural order parameter, defined as \(\bar{\Psi}_{12}(i)\), for each particle \(i\).
A clear correlation emerges---regions of  high structural order (dark areas in Fig.~\ref{fig:corr-qc}\ref{fig:qc-structure}) consistently correspond to zones of low dynamic propensity, while structurally disordered regions  (bright areas) align with  enhanced dynamic mobility. This correlation mirrors  structure-dynamics relationships observed in glass-forming liquids~\cite{berthierDynamicHeterogeneityAmorphous2011}, yet has not been extensively documented for quasicrystalline phases.
These findings suggest that local symmetry inhibits mobility by stabilizing the quasiperiodic network, whereas structural imperfections—such as phason defects or mismatched tiling domains—facilitate rearrangements. Thus a structure-dynamics  correlation, long recognized in supercooled liquids, also emerges in quasicrystals, underscoring a broader principle linking structural constraints to dynamical arrest.

To quantify the correlation between dynamical propensity and structural order in  DDQCs, we computed both Spearman's rank correlation coefficient and Kendall's tau~\cite{kendall1990}. Spearman's rank correlation coefficient, denoted by \(\rho_s\), is defined as the Pearson correlation between the ranked variables. Given paired observations $(x_i, y_i)$ for $i = 1, \dots, n$, let $R(x_i)$ and $R(y_i)$ denote the ranks of $x_i$ and $y_i$, respectively. Then,
\begin{equation}
	\rho_s = \frac{\operatorname{cov}(R(x), R(y))}{\sigma_{R(x)} \sigma_{R(y)}},
\end{equation}
where $\operatorname{cov}$ denotes the covariance and $\sigma$ denotes the standard deviation of the ranked variables.
In the following sections we report the absolute value, \(\left\lvert \rho_{s} \right\rvert\), as the measure of correlation, since we are focused on a correlation independent of whether it is increasing or decreasing.

Kendall's tau, denoted by $\tau$, measures the ordinal association between two variables based on the number of concordant and discordant pairs. For a dataset of $n$ observations, the coefficient is given by
\begin{equation}
	\tau = \frac{(C - D)}{\binom{n}{2}},
\end{equation}
where $C$ is the number of concordant pairs, $D$ is the number of discordant pairs, and $\binom{n}{2}$ is the total number of pairwise comparisons.
Similarly as with Spearman's \(\rho_{s}\), we report the absolute value for Kendall's tau \(\left| \tau \right|\) in the following sections.

Using these measures, we find a strong negative correlation between dynamic propensity and structural order in the DDQC, with  \(\left| \rho_s \right| \approx 0.59\) and \(\left| \tau \right| \approx 0.41\) These results confirm an tight relationship between structural order and dynamic mobility in the DDQC.

Additionally, we identify discrete pockets of localized structural disorder that  correspond precisely to small-scale collective particle rearrangements. These disordered regions frequently appear at the boundaries between different tiling arrangements and at phason defects, consistent with the particle flip dynamics previously described for QCs~\cite{engelDynamicsParticleFlips2010}. Within these localized regions, tile reorientation and restructuring occur through correlated particle motions, while the overall tile composition of the DDQC remains invariant. 

Identifying  this structure-dynamics relationship provides important insights into the microscopic mechanisms that facilitate particle mobility in QCs. Despite their long-range order, these results suggest that QCs share key dynamical features akin to those observed in  amorphous systems.

\begin{figure*}
	\includegraphics[width=0.95\textwidth]{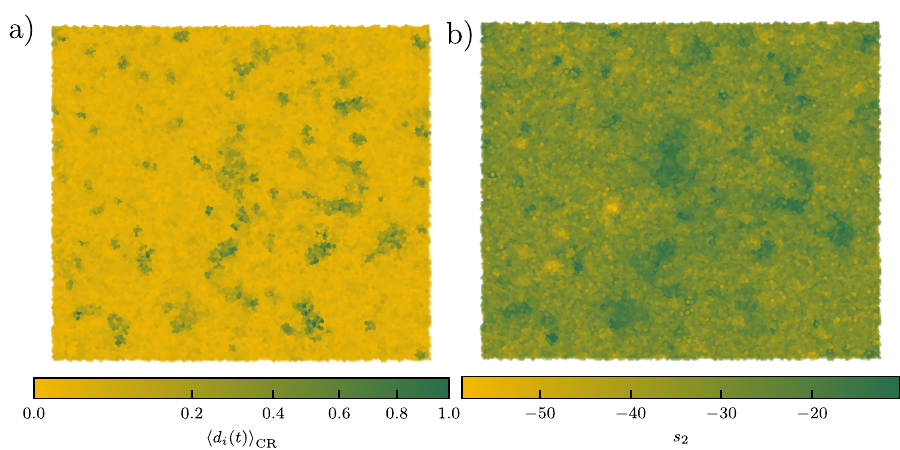}
	\caption[corr-qc]{
		\begin{enumerate*}[label=\textbf{\alph*})]
            \item Dynamic propensity with cage-relative coordinates $\langle d_{i}(t)\rangle_{\mathrm{CR}}$ and \label{fig:corr-glass-propensity}
            \item local structural order parameter \(s_{2}(i)\) for a   supercooled binary liquid mixture with \(N = 16384\) particles at temperature \(k_{B} T / \epsilon = 0.95\) and density \(\rho \sigma^{2} = 0.75\), evaluated over a time interval  \(t = 10^{6} \tau_{G}.\) We use \(M=50\) replicas of the same initial configuration with velocities randomly sampled  from a Maxwell-Boltzmann distribution, to compute the average  absolute displacement across all replicas.
			\label{fig:corr-glass-structure}
		\end{enumerate*}
	}
	\label{fig:corr-glass}
\end{figure*}

\subsection{Correlation Between Dynamic Propensity and Local Order in Supercooled Binary Liquid Mixtures}
We now turn our attention to the supercooled liquid,  applying the same  analysis to examine the link between local structure and particle mobility. 
To quantify particle mobility, we calculate the dynamic propensity of each particle using the same protocol as for the DDQC. For  a fixed initial configuration, we performed \(N=50\) independent simulations with velocities randomly sampled from a Maxwell-Boltzmann distribution and averaged the absolute displacement of each particle over these runs. This approach isolates the structural contribution to dynamical heterogeneity by averaging out kinetic effects. 

The resulting dynamic propensity map for a supercooled binary liquid mixture with $N=16384$ particles at temperature $k_BT/\epsilon=0.95$ and density $\rho\sigma^2=0.75$ is shown in Fig.~\ref{fig:corr-glass}a. 
At this temperature, the system exhibits pronounced dynamical heterogeneity, with distinct regions of high and low mobility that persist over the analyzed time interval. This heterogeneity aligns  with the peaks observed in the non-Gaussian parameter (Fig.~\ref{fig:alfa2}b), confirming that particles in the supercooled liquid experience  different local structural environments that affect their mobility.

In this system, bond-orientational symmetry is not a suitable measure of structural order due to the absence of long-range correlations. Instead, we use the local two-body excess entropy, \(S_{2}(i)\), which captures short-range translational order and has been shown to correlate strongly with dynamics in disordered systems. The two-body excess entropy \(S_{2}(i)\) for particle \(i\) is defined as
\begin{eqnarray}
	S_{2}(i) &=& \nonumber \\
	& & \hspace{-15mm} -\frac{1}{2} \sum_{\beta} \rho_{\beta} \int d \bm{r} \left[g^i_{\alpha\beta}(\bm{r})\ln g^i_{\alpha\beta}(\bm{r}) - g^i_{\alpha\beta}(\bm{r}) + 1\right],
	\label{eq:S2}
\end{eqnarray}
where \(\alpha \in \{S,L\}\) denotes the species of the central particle \(i\), \(\beta \in \{S,L\}\) indicates the species of the surrounding particles in the binary mixture, with $S$ and $L$ referring to the small and large species, respectively. Here, \(\rho_{\beta}\) is the number density of species \(\beta\), and \(g^i_{\alpha\beta}(\bm{r})\) is the pair correlation function between  central particle $i$ of type \(\alpha\) and surrounding particles of type \(\beta\). Physically, \(S_{2}\) quantifies the information content associated with local structural correlations in  particle arrangements, with more negative values indicating higher local structural order. To reduce statistical noise, we compute the time-averaged structural order parameter \(s_{2}(i)\) by averaging $S_2(i)$ over a period of \(10^{6}\tau_{G}\), consistent with the timescale used for the orientational order parameters in the DDQC analysis.

We present the structural order parameter map \(s_{2}(i)\) in Fig.~\ref{fig:corr-glass}b, which reveals a strong  correlation with dynamic propensity. Regions of high structural order (brighter areas in the figure, corresponding to more negative values of \(s_{2}(i)\)) consistently align with zones of reduced mobility, while regions of low structural order (darker areas) coincide with zones of enhanced particle motion. This pattern indicates that local structural arrangements strongly influence particle dynamics, with more ordered regions inhibiting mobility and less ordered regions facilitating it.

We quantify the correlation between dynamic propensity and structural order using Spearman's rank correlation coefficient and Kendall's tau, obtaining  \(\left| \rho_s \right| \approx 0.28\) and \(\left| \tau \right| \approx 0.19\), respectively. These values indicate a moderate positive correlation, confirming that regions with lower structural order correspond to higher particle mobility  in the supercooled liquid.

\begin{figure}
	\includegraphics[width=0.7\columnwidth]{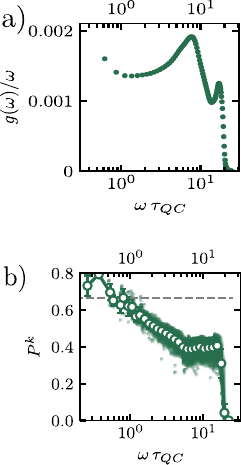}
	\caption[vdos-qc]{
		\begin{enumerate*}[label=\textbf{\alph*})]
            \item Reduced vibrational density of states \(g(\omega) / \omega\) and \label{fig:vdosqc-img}
            \item participation ratio \(P^{k}\) as a function of frequency $\omega \tau_{QC}$ for a dodecagonal quasicrystal with \(N=10976\) particles, calculated for a configuration equilibrated at parent temperature \(k_{B}T/\epsilon = 0.17\) and subsequently energy-minimized.
			Circles denote binned averages to clarify the overall trend.\label{fig:vdos-qc-pr}
		\end{enumerate*}
	}
	\label{fig:vdos-qc}
\end{figure}

This  correlation between structural order and dynamics in the supercooled liquid closely mirrors our findings for the DDQC, despite the fundamentally different structural organizations of the two systems. In both cases, particles residing in locally ordered environments exhibit reduced mobility, suggesting a common underlying principle: local structural order, regardless of its specific form, constrains particle motion in similar ways across diverse condensed matter systems. While this structure-dynamics relationship is well-established in glass-forming liquids~\cite{tongRevealingHiddenStructural2018,paretAssessingStructuralHeterogeneity2020a,richardPredictingPlasticityDisordered2020}, its emergence in quasicrystalline dynamics highlights a novel and potentially unifying connection.

Nevertheless, it is important to recognize that this structure-dynamics  relationship is not unique to amorphous and quasicrystalline systems. Similar patterns of dynamical heterogeneity have been observed in crystalline materials containing defects~\cite{meerDynamicalHeterogeneitiesDefects2015a}, where the reduced mobility in well-ordered regions and enhanced mobility near defects generate comparable spatial variations in dynamics. This observation underscores that, while the structural-dynamic correlations   in the DDQC and supercooled liquid appear qualitatively similar, the microscopic mechanisms driving these patterns may be fundamentally different. A more detailed analysis of vibrational modes and single-particle trajectories is needed to determine whether the underlying  physical processes truly converge or diverge across  these different classes of materials.

To further elucidate the relationship between structure and dynamics in these systems, we turn to their vibrational properties. By analyzing the vibrational density of states, mode localization, and associated topological features, we aim to reveal how structural organization is reflected in the normal mode spectrum, and to determine whether the similarities observed in dynamics extend into the frequency domain.

\section{\label{sec:vibrational} Vibrational properties}

\subsection{Vibrational density of states}
Having established a clear correlation between structure and dynamics, we now shift our focus to the frequency domain and investigate how structural order appears in the vibrational modes of the system. 
To analyze these properties, we compute the dynamical matrix for each system, defined as~\cite{vaibhavExperimentalIdentificationTopological2025}
\begin{equation}
	H_{ij} = \frac{1}{\sqrt{m_{i} m_{j}}} \frac{\partial^{2} U}{\partial \bm{r}_{i} \partial \bm{r}_{j}} \, ,
	\label{eq:hessian}
\end{equation}
where \(\bm{r}_{i}\) represents the spatial coordinate of particle \(i\), and \(m_{i}\) denotes its mass. The potential energy of the system is given by \(U = \sum_{i < j} u_{ij}(r)\), where \(u_{ij}(r)\) denotes the interaction potential between particles \(i\) and \(j\).

For both systems, we compute the dynamical matrix \(H_{ij}\) at an energy-minimized configuration. These configurations were first equilibrated at specific parent temperatures: \(k_{B}T/\epsilon = 0.17\) for the DDQC and \(k_{B}T/\epsilon = 0.1\) for the supercooled liquid. The energy-minimization procedure is performed using the Fast Inertial Relaxation Engine (FIRE) algorithm~\cite{bitzek2006structural}, which effectively quenches the systems to zero temperature, ensuring they reside at local minima of their respective potential-energy surfaces. The choice of the parent temperature is crucial  because it determines which region of the  energy landscape  each system explores before minimization. After constructing the Hessian matrices, we diagonalize them to obtain the full spectrum of eigenvalues and eigenvectors.

From the eigenvalues \(\lambda_l\), with \(l = 1, 2, \dots, 2N\) and \(N\) the total number of particles, we obtain the corresponding normal mode frequencies as \(\omega_l = \sqrt{\lambda_l}\). Using these frequencies, we compute the vibrational density of states, \(g(\omega)\), defined as
\begin{equation}
	g(\omega) = \frac{1}{2N - 2} \sum_{k} \delta(\omega - \omega_{k}) \, .
	\label{eq:vdos}
\end{equation}
To highlight the distribution of low-frequency modes, we also consider the reduced vibrational density of states,
\(g(\omega)/\omega\).
The vibrational density of states (vDOS) counts the number of vibrational modes at each frequency and forms the basis for calculating various thermodynamic properties, such as the heat capacity~\cite{zeller1971thermal,ramos2022low}.
In its reduced form, \(g(\omega)/\omega\) provides additional insight into the localization and excess of low-frequency vibrational modes.
To compute the vDOS, we bin the frequencies into uniform intervals, count the number of modes within each bin, and normalize by the total number of modes and the bin width.
To obtain a smooth representation of the vDOS, we  apply a Gaussian kernel density estimation to the binned data, effectively averaging over neighbouring bins to reduce statistical noise.

We present the vDOS for the DDQC in Fig.~\ref{fig:vdos-qc}\ref{fig:vdosqc-img}.
In two-dimensional systems, the Debye model predicts that the vDOS scales linearly with frequency at low \(\omega\), i.e. \(g(\omega) \propto \omega\),  reflecting   contributions from long-wavelength acoustic modes.
Consistent with this, the reduced vDOS \( g(\omega) / \omega \), remains  roughly constant at low frequencies, aside from the first three modes (Goldstone modes) associated with the translational degrees of freedom, which appear as slight deviations from the constant scaling expected from the Debye model in the initial data points in Fig.~\ref{fig:vdos-qc}\ref{fig:vdosqc-img}. 
At intermediate frequencies (\(\omega \approx 8\)), we observe a pronounced peak reminiscent of the boson peak observed in amorphous solids, though distinct in origin.
In the DDQC, this peak likely arises from an excess of vibrational modes associated with phason dynamics, whic constitute low-energy rearrangements inherent to the quasiperiodic order.
Similar features have been reported in experimental studies of icosahedral QCs~\cite{brandVibrationalDensityStates2000,brandDynamicsIcosahedralQuasicrystal2001} and in theoretical investigations in two dimensions~\cite{odagakiElectronicVibrationalSpectra1986}.
At higher frequencies (\(\omega > 10\)), we identify a secondary peak  following the initial decay of the vDOS, which we attribute to localized vibrational modes.
These high-frequency modes likely correspond to collective particle motions that contribute to the diffusive dynamics of QCs.

Figure~\ref{fig:vdos-glass}\ref{fig:vdos-glass-vdos} displays the normalized vDOS for the supercooled liquid. 
As in the DDQC, the three lowest frequencies correspond to the Goldstone modes, which have higher values in the reduced vDOS due to the eigenvalues being close to zero.
Beyond these three modes, the reduced vDOS, \(g(\omega)/\omega\), remains approximately constant at low frequencies, consistent with the Debye scaling law  for two-dimensional systems. This  indicates that the low-frequency vibrational modes in the supercooled liquid are predominantly acoustic-like, similar to those in the DDQC.
At intermediate frequencies (\(\omega \approx 10\)), we observe a well-defined peak corresponding to the boson peak, indicative of an excess of vibrational modes beyond  the Debye scaling law prediction. This feature is a hallmark of disordered solids, widely reported in  experiments~\cite{monacoDensityVibrationalStates2006}, simulations~\cite{xuLowfrequencyVibrationalDensity2024}, and theoretical studies~\cite{baggioliVibrationalDensityStates2019}. At higher frequencies,  the vDOS gradually decreases, marking a transition from extended acoustic-like modes to increasingly localized vibrational excitations.

\subsection{Participation ratio}
We now analyze the spatial character of the normal modes by examining the participation ratio (PR), which quantifies the extent to which vibrational modes are localized or delocalized across the system. The PR for each vibrational mode $k$ is defined as
\begin{equation}
	P^{k} = \frac{{\left( \sum_{i=1}^{N} {\lvert \bm{e}^{i}_{k} \rvert}^{2} \right)}^{2}}{N \left( \sum_{i=1}^{N} {\lvert \bm{e}^{i}_{k} \rvert}^{4} \right)},
	\label{eq:participation-ratio}
\end{equation}
where \(\bm{e}^{i}_{k}\) represents the eigenvector component for particle \(i\) in mode \(k\).
In two-dimensional systems, the participation ratio reaches $P^k = 2/3$ for perfectly extended plane-wave modes, while it approaches $P^k \approx 1/N$ for highly localized modes~\cite{laird1991localized}.
This metric  quantifies the effective fraction of particles that significantly participate to each vibrational mode, distinguishing between delocalized, collective excitations and those that are spatially confined to small regions of the system.

In Figure~\ref{fig:vdos-qc}\ref{fig:vdos-qc-pr}, we plot the PR spectrum as a function of frequency $\omega$ for the DDQC, calculated from the energy-minimized configuration at parent temperature $k_BT/\epsilon = 0.17$. 
At low frequencies ($\omega \tau_{QC} < 2)$, the $P^k$ values approaches  the theoretical plane-wave limit of $2/3$. 
As frequency increases into the intermediate range ($2 < \omega \tau_{QC} < 20$), the $P^k$ values decrease with increasing frequency as the vibrational modes become more localiezed.   At higher frequencies ($\omega > 20$), the $P^k$ sharply  drops to zero. 

\begin{figure}
	\includegraphics[width=0.7\columnwidth]{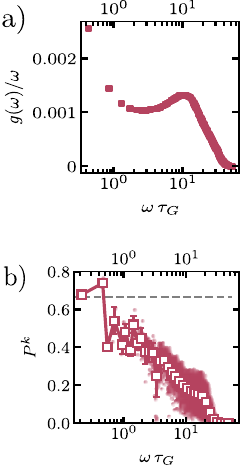}
	\caption[vdos-g]{
		\begin{enumerate*}[label=\textbf{\alph*})]
            \item Reduced vibrational density of states \(g(\omega) / \omega\) and \label{fig:vdos-glass-vdos}
            \item participation ratio \(P^{k}\) as a function of frequency  $\omega \tau_G$, calculated for a configuration equilibrated at parent temperature \(k_{B}T/\epsilon = 0.1\) and subsequently energy-minimized.
              Squares denote binned averages to clarify  the overall trend.\label{fig:vdos-glass-pr}
		\end{enumerate*}
	}
	\label{fig:vdos-glass}
\end{figure}

Figure~\ref{fig:vdos-glass}\ref{fig:vdos-glass-pr} presents the PR spectrum for the supercooled liquid, calculated from the energy-minimized configuration at parent temperature $k_BT/\epsilon = 0.1$. 
At low frequencies ($\omega \tau_G < 1$), the PR approaches $0.60$, slightly below the theoretical plane-wave limit of $2/3$, which reflects the system's inherent disorder even in its most extended vibrational modes.
As frequency increases into the intermediate range ($1 < \omega \tau_G< 10$), we observe substantial variability in the PR values, with the standard deviation reaching approximately $0.15$ at $\omega \tau_G\approx 5$.
This suggests that modes at similar frequencies can have vastly different spatial distributions, with some being more localized than others.
This heterogeneity in mode localization mirrors the dynamical heterogeneity observed in the non-Gaussian parameter analysis (Section~\ref{sec:alfa2}), indicating a potential correlation between vibrational mode localization  and the propensity for particle rearrangements.
At higher frequencies ($\omega \tau_G> 10$), the $P^k$ values drop below $0.2$, signifying highly localized modes confined to small clusters of particles.
The binned data points (squares with error bars) are calculated using uniformly spaced frequency windows, containing at least 15 modes each, with error bars showing one standard deviation.
These fluctuations are  physically meaningful and reflect real variations in mode character within each frequency band rather than statistical noise.

The PR spectrum of the DDQC reveals both similarities and notable differences  compared to  the supercooled liquid.
Both systems exhibit the general trend of decreasing $P^k$ with increasing frequency $\omega$, but the DDQC maintains $P^k$ values consistently closer to the plane-wave limit ($\sim 0.65$) over a broader frequency range ($\omega \tau_{QC}< 2$) than the supercooled liquid ($\omega  \tau_{G}< 1$).
This quantifiable difference ($\sim 8\%$ higher $P^k$ at $\omega  \tau_{QC/G}= 1.5$) reflects the more coherent structural organization of the DDQC despite its aperiodicity.
At intermediate frequencies, both systems show substantial fluctuations in PR, but the DDQC's average $P^k$ remains slightly higher, indicating that its vibrational modes are more extended.
This enhanced collectivity likely arises from the DDQC's quasi-long-range order, which creates more consistent local environments than those in the supercooled liquid. 
These PR characteristics increase our understanding of the systems' dynamics. In the DDQC, the more extended vibrational modes  at intermediate frequencies align with  its collective rearrangement mechanisms, such as phason flips, while the supercooled liquid's more localized modes  reflect its more heterogeneous particle mobility. This connection between vibrational localization and dynamical behavior provides a natural bridge to our topological analysis below, where we  examine how these vibrational properties are reflected in the spatial distribution of topological defects within the eigenvector fields.

\subsection{Topological defects}
\begin{figure}
	\includegraphics[width=0.8\columnwidth]{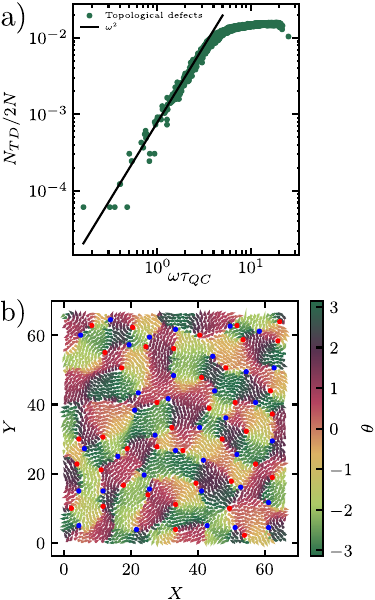}
	\caption[defects-qc]{
	\begin{enumerate*}[label=\textbf{\alph*})]
        \item Frequency dependence of topological defect density \(N_{TD}/2N\), with theoretical acoustic scaling \(N_{TD}/2N \propto \omega^{2}\) shown in black.\label{fig:qc-distribution}
        \item Eigenfield visualization at frequency \(\omega \tau_{QC} = 2.463\), showing topological defects with charge \(q = +1\) (red circles) and \(q = -1\) (blue circles). Arrows represent the normalized eigenvector components.\label{fig:qc-eigenfield}
		\end{enumerate*}
	}
	\label{fig:defects-qc}
\end{figure}

We now analyze the topological characteristics of the vibrational eigenmodes, which provide deeper insight into the fundamental differences between the DDQC and the supercooled liquid beyond their dynamical similarities. Following the approach of Ref.~\cite{wuTopologyVibrationalModes2023}, we identify the  topological defects in the eigenvector fields and examine how their spatial distribution varies with frequency.

For each vibrational mode with eigenvector field \((\bm{e}_{i}^{x}, \bm{e}_{i}^{y})\) at frequency \(\omega_{i}\), we construct a continuous angle field \(\theta(\bm{r})\) on a \(120 \times 120\) spatial grid, representing the local orientation of the displacement vectors
\begin{equation}
	\tan{\theta(\bm{r})} = \frac{\sum_{i} w(\bm{r} - \bm{r}_{i}) \bm{e}_{i}^{y}}{\sum_{i} w(\bm{r} - \bm{r}_{i}) \bm{e}_{i}^{x}}
	\label{eq:theta}
\end{equation}

where \(w(\bm{r} - \bm{r}_{i}) = \exp(-|\bm{r} - \bm{r}_{i}|^{2}/\sigma^{2})\) with \(\sigma \)  a Gaussian weight function that smooths the discrete eigenvector field~\cite{goldenberg2005friction,oyama2019avalanche}. Using this continuous representation, we then identify topological defects by calculating the winding number around each elementary cell
\begin{equation}
	q = \frac{1}{2\pi} \oint \nabla \theta \cdot d\bm{l},
	\label{eq:charge}
\end{equation}
which equals zero in defect-free regions and takes values \(q = \pm 1\) at topological singularities.

Figure~\ref{fig:defects-qc} presents results for the DDQC equilibrated at \(k_{B}T/\epsilon = 0.17\) and subsequently energy-minimized. The eigenvector field at \(\omega \tau_{QC}= 2.463\), shown in Fig.~\ref{fig:defects-qc}b,  reveals numerous topological defects in the form of vortex-like structures, where the orientation of eigenvectors winds around singular points. These defects typically appear in pairs of opposite charge, consistent with topological charge conservation.

\begin{figure}
	\includegraphics[width=0.75\columnwidth]{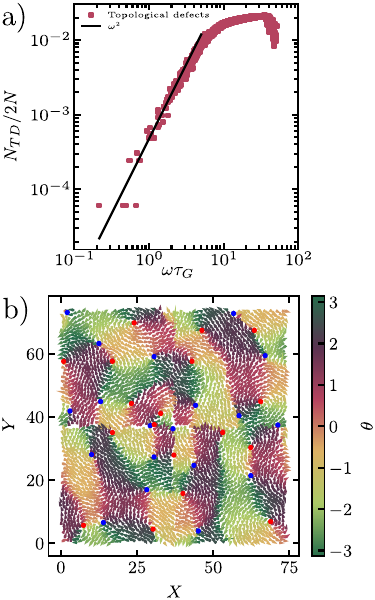}
	\caption[defects-glass]{
		\begin{enumerate*}[label=\textbf{\alph*})]
            \item Frequency dependence of topological defect density \(N_{TD}/2N\). The black line shows the theoretical scaling \(N_{TD}/2N \propto \omega^{2}\) expected for acoustic modes. \label{fig:glass-distribution}
            \item Eigenfield visualization at frequency \(\omega \tau_{G} = 2.397\), showing topological defects with charge \(q = +1\) (red circles) and \(q = -1\) (blue circles). Arrows represent the normalized eigenvector components.\label{fig:glass-eigenfield}
		\end{enumerate*}
	}
	\label{fig:defects-glass}
\end{figure}

In Figure~\ref{fig:defects-glass}, we present results for the supercooled liquid, equilibrated at \(k_{B}T/\epsilon = 0.1\) and energy-minimized. While, the eigenfield at \(\omega \tau_G= 2.397\) in Fig.~\ref{fig:defects-glass}b  exhibits qualitatively similar eigenvector patterns and  vortex-like topological defects as observed for the DDQC, the frequency dependence of the density of topological defects  \(N_{TD}/2N\) reveals clear distinctions between the two systems as shown in Fig.~\ref{fig:defects-qc}a and Fig.~\ref{fig:defects-glass}a.

At low frequencies, the density of topological defects \(N_{TD}/2N\) in both systems  follows the predicted \(\omega^{2}\) scaling law~\cite{wuTopologyVibrationalModes2023}, consistent with  the acoustic properties of extended vibrational modes. This confirms that both the supercooled liquid and the DDQC support proper long-wavelength acoustic modes  despite their structural differences. At higher frequencies, the defect density in the supercooled liquid (Fig.~\ref{fig:defects-glass}a) decreases sharply, reflecting increasing mode localization and  defect annihilation. In contrast, the DDQC maintains a  significantly higher  defect density (Fig.~\ref{fig:defects-qc}), with a more gradual crossover and much slower decay.

This persistence of topological defects in the DDQC's high-frequency modes highlights the impact of its unique structural organization. In the supercooled liquid, the absence of long-range order leads to strong scattering and rapid localization of high-frequency modes, resulting in the quick annihilation of topological defects. In contrast, the DDQC's quasi-long-range order supports more coherent pathways for vibrational energy propagation.  As a result, topological defects remain prevalent even at higher  frequencies, reflecting the sustained complexity of the eigenfields. The slower rate of defect annihilation in the DDQC indicates that its vibrational modes retain an  extended character over a wider frequency range, consistent with our participation ratio analysis in Section~\ref{sec:vibrational}.

The distinct behavior of topological defects at high frequencies provides a clear signature of the fundamental differences between the two systems.  Although the DDQC and the supercooled liquid exhibit similar dynamical heterogeneity and structure-dynamics correlations, they diverge sharply in their vibrational responses—a direct consequence of the quasi-long-range order in the DDQC versus the complete structural disorder in the supercooled liquid. This contrast underscores the value of topological defect analysis: by probing the spatial organization of vibrational modes, it becomes possible to distinguish between structural classes that may otherwise appear dynamically similar.

\section{\label{sec:crystal-similarities} Dynamical and vibrational similarities to a square crystal}

The dodecagonal quasicrystal (DDQC) and the supercooled liquid exhibit remarkably similar dynamical behavior despite their contrasting structural organizations, while their vibrational properties reveal pronounced differences. To further contextualize these findings, we  introduce a square crystal as a structurally ordered reference system. By comparing these three phases, crystalline, quasiperiodic, and amorphous, we aim to disentangle which aspects of dynamics and vibrations are universal, and which are uniquely shaped by disorder or quasiperiodicity.

We simulated a two-dimensional square crystal composed of \(N = 4096\) particles at a fixed density of \(\rho \sigma^{2} = 0.85\), using the same pairwise interaction potential employed for the DDQC.
To ensure consistency with our previous analyses, we applied the same linear cooling protocol as for the DDQC, followed by equilibration across a temperature range \(k_{B} T / \epsilon \in [0.12, 0.2]\).
Vibrational properties were computed by minimizing the potential energy using the FIRE algorithm and analyzing the dynamical matrix, as performed for the quasicrystal and supercooled liquid.
Dynamical observables--including the mean-squared displacement and non-Gaussian parameter--were evaluated from microcanonical (\(NVE\)) simulations, ensuring that all three systems were compared using the same numerical framework.
\begin{figure}
    \includegraphics[width=0.9\columnwidth]{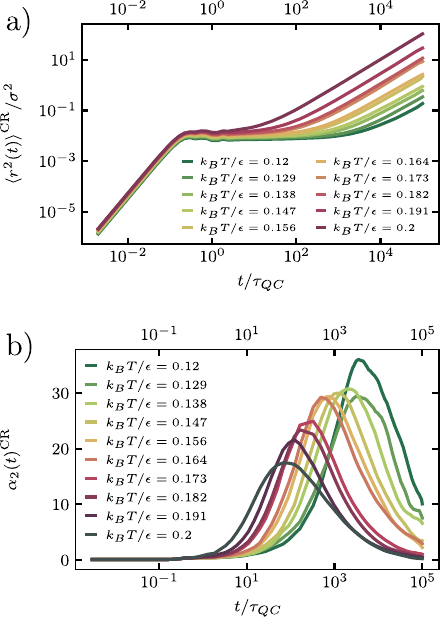}
	\caption[dynamics-square]{
		\begin{enumerate*}[label=\textbf{\alph*})]
            \item Mean-squared displacement \(\langle \Delta r^{2}(t) \rangle_{\textrm{CR}}\) and \label{fig:msd-square}
            \item non-Gaussian parameter \(\alpha_{2}(t)\) as a function of time \(t\) for the two-dimensional square crystal at different temperatures.  \label{fig:alfa2-square}
        \end{enumerate*}
	}
    \label{fig:dynamics-square}
\end{figure}
The mean-squared displacement (MSD) of the square crystal, shown in Fig.~\ref{fig:dynamics-square}\ref{fig:msd-square}, reveals temperature-dependent dynamics that is similar to, but also diverge from, those seen in the DDQC and supercooled liquid.
At high temperatures, particles exhibit normal diffusive behavior, with the MSD increasing linearly over time. As temperature decreases, a transient plateau emerges in the MSD, signaling the onset of cage-trapping dynamics. However, the plateau is markedly shorter and less developed than in the DDQC or glass, indicating that dynamical arrest is less pronounced in the square crystal. This difference reflects the underlying translational order of the crystal, which limits the configurational freedom necessary for extended trapping. Nonetheless, the presence of defects in the crystal lattice introduces localized dynamical constraints that mimic some aspects of glassy behavior. Prior studies have shown (e.g. Ref.~\cite{meerDynamicalHeterogeneitiesDefects2015a}) that such imperfections can generate short-lived cages and heterogeneous dynamics similar to those in amorphous systems.

The non-Gaussian parameter \(\alpha_{2}(t)\) for the square crystal, plotted in Fig.~\ref{fig:dynamics-square}\ref{fig:alfa2-square}, exhibits pronounced peaks across all temperatures studied, signaling persistent dynamical heterogeneity.
This result reinforces the idea that even in nominally ordered systems, local structural imperfections, such as vacancies, interstitials, or dislocations, can induce heterogeneous particle dynamics akin to those in disordered phases.
The presence of such defects disrupts the uniformity of the lattice and generates localized regions of constrained mobility.
Remarkably, the temporal signatures of \(\alpha_{2}(t)\) in the square crystal are qualitatively similar to those observed in the DDQC and supercooled liquid, suggesting that the emergence of cage-trapping and intermittent motion is not exclusive to disordered or quasiperiodic media.
At the level of dynamical observables alone, distinguishing between crystalline, quasiperiodic, and amorphous systems thus proves challenging, at least in two dimensions.

\begin{figure}
	\includegraphics[width=0.75\columnwidth]{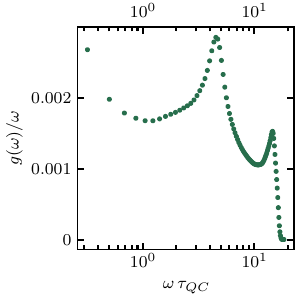}
	\caption[vdos-square]{Reduced vibrational density of states \(g(\omega) / \omega\) of the square crystal as a function of frequency $\omega$, calculated for a configuration equilibrated at parent temperature \(k_{B}T/\epsilon = 0.12\) and subsequently energy-minimized.}
	\label{fig:vdos-square}
\end{figure}

\begin{figure}
	\includegraphics[width=0.75\columnwidth]{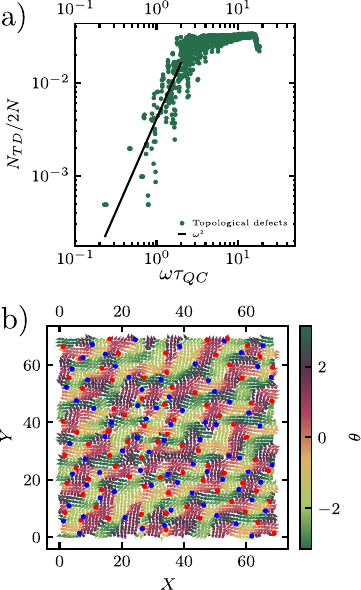}
	\caption[defects-square]{\textbf{Topological defects in vibrational eigenmodes of the square crystal.}
		\begin{enumerate*}[label=\textbf{\alph*})]
            \item Frequency dependence of topological defect density \(N_{TD}/2N\), with theoretical acoustic scaling \(N_{TD}/2N \propto \omega^{2}\) shown in black.\label{fig:square-distribution}
			\item Eigenfield visualization at frequency \(\omega \tau_{QC} = 2.282\), showing topological defects with charge \(q = +1\) (red circles) and \(q = -1\) (blue circles). Arrows represent the normalized eigenvector components.
		\end{enumerate*}
	}
	\label{fig:defects-square}
\end{figure}

In contrast to their dynamical similarity, the vibrational properties of the square crystal differ markedly from those of the DDQC and supercooled liquid.
Figure~\ref{fig:vdos-square} shows the reduced vibrational density of states, \(g(\omega) / \omega\), for the square crystal.
At low frequencies, the vDOS displays the expected Debye  scaling law, reflecting the presence of long-wavelength acoustic modes, similar to the other two systems.
Beyond this regime, however, the vDOS of the square crystal exhibits sharp van Hove singularities—two prominent peaks arising from divergences in the density of states at critical points in the phonon dispersion relations. 
These peaks correspond to the longitudinal and transverse acoustic branches reaching the Brillouin zone boundary and serve as hallmarks of crystalline periodicity. While the DDQC also exhibits secondary peaks, likely associated with phason-related vibrations, they are broader and less sharply defined, indicating a partial suppression of long-range coherence. In contrast, the supercooled liquid shows no such singularities, instead features a broad boson peak at intermediate frequencies that reflects its disordered structure and the localization of high-frequency vibrational modes. Thus, the vDOS provides a clear fingerprint that differentiates crystals from both quasicrystals and glasses.
Finally, we examine the topological defects in the vibrational eigenmodes of the square crystal, shown in Fig.~\ref{fig:defects-square}a.
At low frequencies, the defect density \(N_{TD}/2N\) scales as \(\omega^{2}\), consistent with the acoustic nature of the low-frequency vibrational modes, and as also observed for  both the DDQC and supercooled liquid.
However, as the frequency increases beyond \(\omega\tau_{QC} \sim 1\), the defect density begins to level off. This crossover marks the point where the longitudinal and transverse phonon branches start to mix near  the Brillouin zone boundary, leading to a  breakdown of the ideal plane-wave character of the modes.
Beyond the first van Hove singularity, the crystalline symmetry strongly constraints the possible  eigenvector topologies. The vibrational modes  become increasingly standing-wave-like, preventing the formation of additional topological defects (such as \(q = \pm 1\) vortices). As a result, the defect density saturates at higher frequencies. This behavior contrasts with the DDQC, where quasi-long-range order allows for a wider range of vibrational distortions, maintaining  a higher defect density at elevated frequencies. Thus, the vibrational topology of the square crystal provides a clear distinction from both the DDQC and the supercooled liquid.

The insights gained from comparing crystalline, quasiperiodic, and amorphous systems clarify not only their dynamical and vibrational characteristics but also provide a platform for broader conceptual insights into condensed matter behavior.

\section{\label{sec:conclusions} Concluding Remarks}
In this work, we investigated the dynamical behavior and vibrational properties of two structurally distinct two-dimensional systems: a dodecagonal quasicrystal (DDQC) and a supercooled binary liquid. Despite their differing degrees of structural order, both systems exhibit remarkably similar dynamic features, including cage-trapping plateaus in the mean-squared displacement, pronounced peaks in the non-Gaussian parameter, and strongly heterogeneous particle mobility fields. These findings highlight that complex, glass-like dynamics are not exclusive to disordered systems and that quasiperiodic order can support analogous dynamical signatures. 
By computing dynamic propensity maps and correlating them with local structural order parameters, we demonstrated that local structure strongly influences particle mobility in both systems: regions of high bond-orientational order in the DDQC and regions of highly negative local excess entropy in the supercooled liquid consistently correspond to areas of lower mobility. We thus observe that the structure-dynamics correlation observed for the DDQC closely resembles that in the supercooled liquid. This observation is nontrivial given that such structure-dynamics correlations are widely considered hallmarks of glassy behavior~\cite{berthierDynamicHeterogeneityAmorphous2011}.

Focusing on vibrational properties, we examined in detail the vibrational density of states (vDOS) and participation ratio (PR) for both systems. While both systems exhibit the expected Debye scaling law at low frequencies, their vibrational spectra diverge markedly at higher frequencies. The supercooled liquid exhibits a pronounced boson peak---a hallmark of amorphous solids---whereas the DDQC displays multiple peaks, likely associated with phasonic excitations, which are collective rearrangements unique to quasicrystalline order.
  
To  further analyze the spatial character of  vibrational modes, we examined the distribution of topological defects within the eigenvector fields, following a procedure originally developed for studying plastic events in supercooled liquids \cite{wuTopologyVibrationalModes2023}. This analysis revealed  significant differences in the vibrational responses between the two systems. While both the DDQC and the supercooled liquid exhibit the expected \(\omega^2\) scaling of topological defect density at low frequencies, the DDQC maintains a significantly higher defect density at higher frequencies, with a more gradual crossover. This persistence of defects reflects the influence of quasi-long-range order in the DDQC, which supports coherent vibrations even in frequency regimes where the supercooled liquid exhibits strong localization.

Lastly, we compared the DDQC and supercooled liquid to a square crystal, which serves as a reference point for understanding the structural and dynamical properties of these systems. The square crystal exhibits similar cage-trapping dynamics and dynamical heterogeneity, but its vibrational properties differ, particularly in the presence of well-defined van Hove singularities in the vDOS. This comparison underscores the unique vibrational characteristics of the DDQC and supercooled liquid, which arise from their distinct structural organizations, and truly sets the DDQC in between the supercooled liquid and the square crystal in terms of vibrational properties.

Despite the striking parallels identified in this study, the analogy between glasses and QCs remains incomplete.
Building a more unified framework will likely  require connecting vibrational spectra, elastic responses, and non-affine dynamics across these structurally distinct regimes, an endeavor that could benefit from combined simulation studies and experimental spectroscopic investigations.
A key open question in the glass community concerns the nature of low-frequency vibrational excitations and the observed deviations from the Debye scaling law in the low-frequency vDOS~\cite{lerner2021low,lerner2016statistics,xuLowfrequencyVibrationalDensity2024,lerner2024enumerating}, which we have not fully addressed here.
Additionally, the elastic properties of QCs remain poorly understood, and the link between topological defects and plastic deformations calls for further study.

Recent work showing similar phonon dynamical properties in three-dimensional icosahedral QCs and supercooled liquids~\cite{caoPhononDynamics3D2025} further underscores the deep structural and dynamical connections across these systems in different spatial dimensions.
Extending our analysis to three dimensions could provide new insights, especially by avoiding the Mermin-Wagner fluctuations that we accounted for using the cage-relative analysis in two dimensions.

\begin{acknowledgments}
	E.A.B.-M. thanks Susana Mar{\'i}n-Aguilar, Gerardo de Jes{\'u}s Campos-Villalobos, and Laura Filion for useful discussions. M.D. and E.A.B-M acknowledge funding from the European Research Council (ERC) under the European Union’s Horizon 2020 research and innovation programme (Grant agreement No. ERC-2019-ADG 884902 SoftML).
\end{acknowledgments}

\section*{Data Availability Statement}

The software and codes to generate the data that support the findings of this study are available at the Zenodo repository found in \url{https://doi.org/10.5281/zenodo.16965771} with DOI 10.5281/zenodo.16965771~\cite{bedolla_2025_16965771}.
The repository includes the scripts to generate the data using molecular dynamic simulations, and the scripts to perform the analysis.

\appendix

\section{\label{sec:flips} Single-Particle Trajectories and Local Dynamics}
\begin{figure*}
	\includegraphics[width=0.85\textwidth]{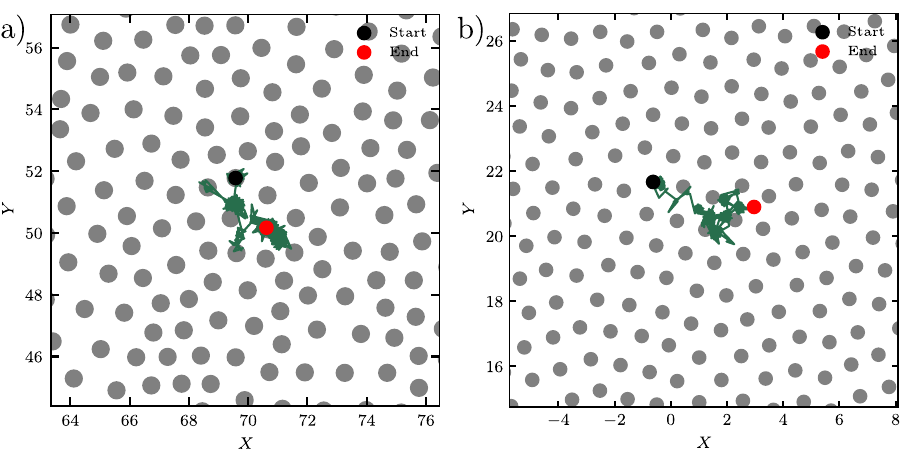}
	\caption[traj]{\textbf{Single-particle trajectories revealing similar intermittent dynamics in structurally distinct systems.} Representative trajectories of individual mobile particles tracked over \(10^{6}\) MD time units. Gray circles show neighboring particles in the initial configuration, with the selected particle's initial position marked by a black circle and final position by a red circle.
		\begin{enumerate*}[label=\textbf{\alph*})]
			\item Trajectory of a mobile particle in the supercooled liquid at \(k_{B} T / \epsilon = 0.95\), showing characteristic cage rattling followed by discrete jumps.
			\item Trajectory of a mobile particle in the dodecagonal quasicrystal at \(k_{B} T / \epsilon = 0.17\), exhibiting similar step-like motion despite the system's quasi-long-range order.
		\end{enumerate*}
	}
	\label{fig:trajectories}
\end{figure*}

In this appendix we show insight into how individual particles navigate their local environments, complementing our analyses from the main sections of mean-squared displacement, non-Gaussian parameters, and vibrational characteristics.

Figure~\ref{fig:trajectories} presents representative trajectories of individual particles in both the DDQC and supercooled liquid, tracked over \(10^{6}\) MD time units. For this analysis, we deliberately selected particles exhibiting significant mobility, as a large fraction of particles in both systems remain effectively immobile during the observation period, a direct manifestation of the dynamical heterogeneity quantified in Section~\ref{sec:alfa2}.
In both cases, particle motion follows a characteristic sequence: (i) initial localized vibrations within a cage formed by neighboring particles, (ii) sudden displacement events enabling particles to escape from their original environment, and (iii) periods of relative stability in new metastable locations before subsequent jumps. This intermittent behavior results in trajectories marked by cage-trapping plateaus of confined motion separated by abrupt transitions.

In the DDQC (Fig.~\ref{fig:trajectories}b), the observed discrete jumps are reminiscent of phason flips—collective rearrangements known to preserve quasicrystalline order while enabling structural relaxation~\cite{engelDynamicsParticleFlips2010,zhaoAtomisticMechanismsDynamics2025}. While our data do not allow for a definitive identification of such events, the qualitative features of the trajectories are consistent with the coordinated motion expected in phason-mediated rearrangements, wherein local tiling configurations are transformed without disrupting the system's overall dodecagonal symmetry. Prior studies have shown that these processes are governed by well-defined activation energies~\cite{kaluginMechanismSelfDiffusionQuasiCrystals1993}, offering a plausible explanation for the thermally activated nature of diffusion in QCs and the temperature dependence of the non-Gaussian parameter observed in Fig.~\ref{fig:alfa2}b.

Interestingly, particles in the supercooled liquid also exhibit qualitatively similar trajectory patterns (Fig.~\ref{fig:trajectories}a), despite the absence of underlying quasiperiodic order. Following extended periods of cage rattling, particles occasionally undergo abrupt displacements that suggest cooperative rearrangements involving multiple neighbors. While our data do not resolve specific microscopic mechanisms, the observed events resemble structural transitions described in the glass literature, such as T1-like processes~\cite{eckmann2008ergodicity,zhou2015t1,ishinoMicroscopicStructuralOrigin2025}, which involve neighbor exchanges and collective particle motion. These processes have been proposed as order-preserving transitions that maintain local structural correlations, drawing a conceptual parallel to the phason flips in quasicrystals,  which enable structural rearrangement without disrupting global order.

The similarity in single-particle trajectories provides a microscopic foundation for the comparable features observed in our analyses of the mean-squared displacement (Section~\ref{sec:msd}) and non-Gaussian parameters (Section~\ref{sec:alfa2}). In both systems, the cage-trapping plateaus in the MSD reflect particle confinement within transient cages formed by neighboring particles, while eventual cage-breaking events lead to diffusive behavior at longer time scales. The spatial and temporal heterogeneity of these cage-breaking events  manifests as pronounced peaks in the non-Gaussian parameters.

These observations also elucidate the connection between the structure-dynamics correlations identified in Section~\ref{sec:dynamics} and the distinct vibrational properties analyzed in Section~\ref{sec:vibrational}. In both systems, regions of higher structural order exhibit reduced mobility, as their locally stable configurations suppress the structural rearrangements necessary for particle diffusion. However, the mechanisms that facilitate these rearrangements differ fundamentally: in the DDQC, they are governed by phason dynamics embedded in a quasiperiodically ordered framework, whereas in the supercooled liquid, they arise from cooperative motions within a disordered landscape.

This mechanistic difference is most clearly reflected in the topological defect analysis of vibrational modes: the DDQC sustains a higher defect density at high frequencies, a consequence of its quasi-long-range order that enables coherent vibrations even at shorter wavelengths. The analysis of single-particle trajectories thus completes our comparative analysis by connecting the macroscopic similarities in dynamical heterogeneity to the microscopic differences in structural organization and vibrational behavior.

\bibliography{bibliography}

\end{document}